\documentclass[11pt]{article}
\usepackage{amssymb,verbatim,epsfig}
\usepackage{amsmath,setspace}
\usepackage{multicol}
\usepackage{graphicx}
\usepackage{sidecap}
\usepackage{wrapfig}
\usepackage{url}
\usepackage[comma,sort&compress,square]{natbib}
\usepackage[english]{babel}
\usepackage{amssymb, amsmath, verbatim, epsfig}
\usepackage{setspace}

\newcommand{\eq}[1]{Eq. (\ref{#1})}
\newcommand{\fig}[1]{{\bf Fig. \ref{#1}}}
\newcommand{\tab}[1]{Table \ref{#1}}
\newcommand{\et}{{\it et al}}

\spacing{1}    
\usepackage[hang,small,bf]{caption}

\title{\bf 
Thermocapillary-driven fluid flow within microchannels} 
\author{Guillermo J. Amador, Ahmet F. Tabak, Ziyu Ren, Yunus Alapan, Oncay Yasa, Metin Sitti$^{*}$ \\
Physical Intelligence Department, Max Planck Institute for Intelligent Systems, \\
Stuttgart, Germany \\
$^*$Correspondence to: sitti@is.mpg.de
}

\begin{document}
\maketitle 

\begin{abstract}
	Surface tension gradients induce Marangoni flow, which may be exploited for fluid transport. At the micrometer scale, these surface-driven flows can be more significant than those driven by pressure. By introducing fluid-fluid interfaces on the walls of microfluidic channels, we use surface tension gradients to drive bulk fluid flows. The gradients are specifically induced through thermal energy, exploiting the temperature dependence of a fluid-fluid interface to generate thermocapillary flow. In this report, we provide the design concept for a biocompatible, thermocapillary microchannel capable of being powered by solar irradiation. Using temperature gradients on the order of degrees Celsius per centimeter, we achieve fluid velocities on the order of millimeters per second. Following experimental observations, fluid dynamic models, and numerical simulation, we find that the fluid velocity is linearly proportional to the provided temperature gradient, enabling full control of the fluid flow within the microchannels.
 \\ %144 words

%This platform is capable of continuously driving fluid flows without the use of conventional fluid pumps and irrespective of the electrical conductivity of the working fluid.

\end{abstract}

\section{Introduction}

%At small length scales, capillary forces become prevalent due to their linear scaling with length, while long-range body forces scale with area or volume.

Microfluidic devices have greatly improved within the last 30 years following the development of microelectromechanical systems (MEMS) \cite{Fan,Ho}. These miniature devices manipulate small volumes of fluids, on the order of nano and microliters, for fundamental studies of individual cells and macromolecules as well as clinical assays \cite{Stone}. At these length scales, typically less than a millimeter, viscous and other surface forces dominate, providing challenges for controlling fluid transfer \cite{Sitti}. Traditionally, mechanical pumps are utilized for generating the pressure gradients necessary for driving fluid flow. However, other attractive methods have arisen due to their efficiency at small scales and controllability given certain signals, including the use of enzymatic reactions \cite{Sengupta,Das}, capillary effects, sound, and electric and magnetic fields \cite{Stone}.  

Surface tension arises from an imbalance of intermolecular forces between molecules on the surface of a fluid. At small length scales, these forces are prevalent. Therefore, fluid transport via capillary action is very attractive for applications at the micrometer scale \cite{Stone,Darhuber,Karbalaei}. Biological systems exploit such surface forces \cite{Bourouiba}, either for propulsion \cite{Bush}, spore dispersal \cite{Money}, adhesion \cite{Federle}, feeding \cite{Kim}, or respiration \cite{Notter}. These biological systems use changes in surface tension to generate large forces at small length scales.

Fluid motion due to surface tension gradients has been documented for over 150 years \cite{Thomson,Marangoni}. Gradients in surface tension can be produced magnetically, electrically, chemically, or thermally. In engineered systems, fluid transport using electric fields has been widely applied \cite{Prins,Janocha,Chung,Luo}. By using electrically conductive fluids, applied electric potentials will alter the apparent surface tension at a liquid-solid interface--commonly referred to as electrowetting \cite{Colgate}. The use of electric fields to induce surface tension gradients and drive fluid flows is especially appealing for use in electronics, such as for liquid-cooling systems in microelectronics. However, it is limited to electrically conductive fluids.

Temperature differences also induce surface tension gradients \cite{Benard}, since the interfacial tension of fluid-fluid interfaces depends on the local temperature \cite{Shereshefsky}. While this phenomenon has been studied for over 100 years, it has only recently received attention for use in practical applications \cite{Kataoka,Maggi,Sammarco,Ohta,Hu}. In this study, we demonstrate the first enclosed thermocapillary-driven pump for microfluidics. This pump was proposed theoretically in a previous study \cite{Baier}. Here, we present the proof of concept microfluidic channel with the walls lined with fluid-filled cavities that introduce fluid-fluid interfaces (\fig{Fig1}a-c). A temperature gradient across the interfaces was introduced using external heating and/or cooling. By inducing temperature gradients on the order of degrees Celsius per centimeter, we pumped water with flow speeds on the order of millimeters per second.

\section{Results} 

%\subsection{Marangoni flow within microfluidic channels}

\begin{figure}
\begin{center}
\includegraphics[width=175mm, keepaspectratio=true]
      {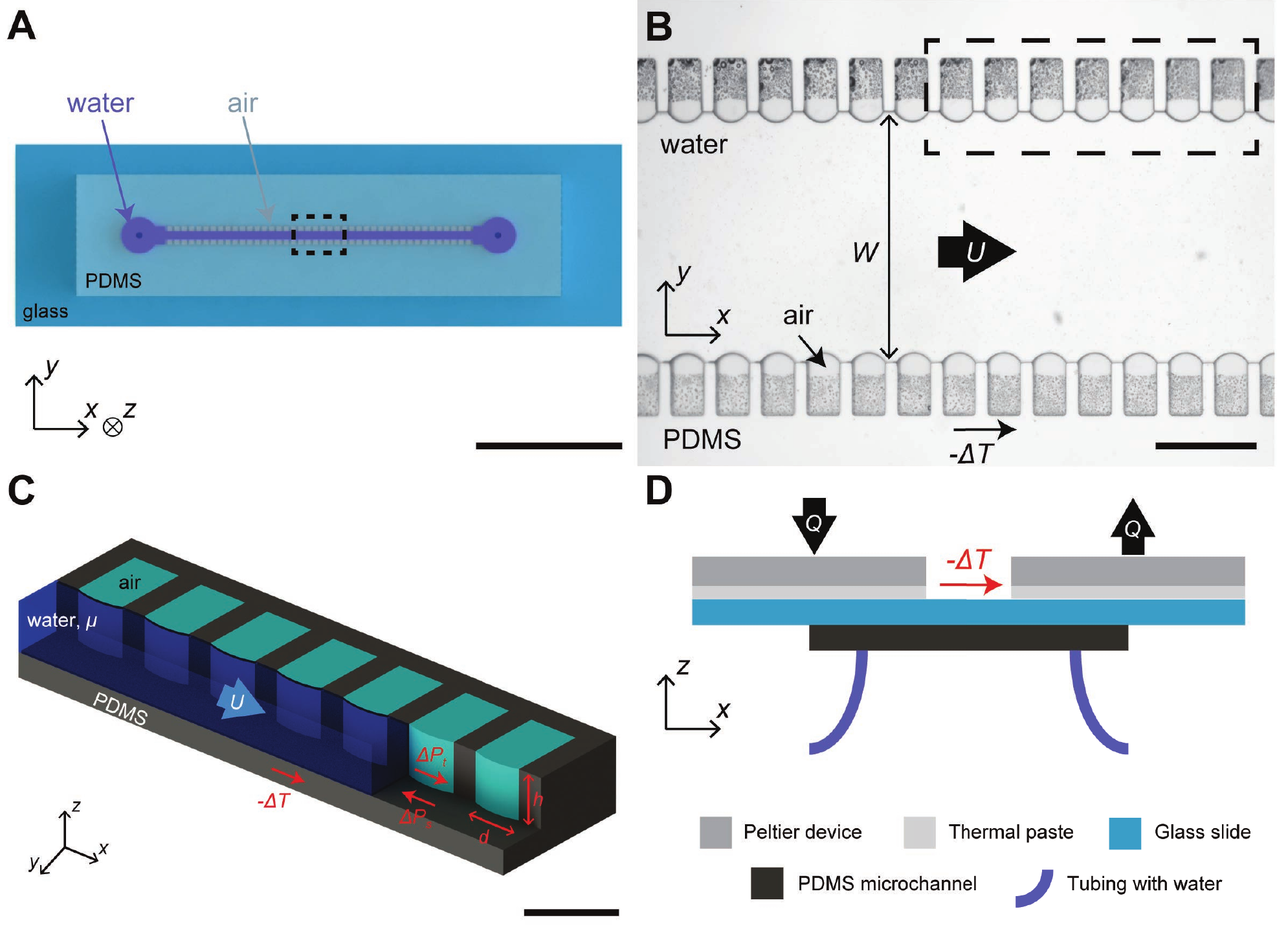}
      \end{center}
      \vspace{0in}
      \caption{{\bf Design concept for thermocapillary-driven microfluidic channels.} (a) Schematic of the entire microfluidic channel with air pockets trapped along the side walls. The dashed rectangle represents the region depicted in (b). (b) Microfluidic channel filled with water and air trapped all along the wall. The dashed rectangle represents the region depicted in the schematic in (c). (c) Schematic of the inside of a microfluidic channel with water (blue) and air pockets (cyan). The dimensions of the cavities are labeled, as well as the two competing pressure gradients arising from temperature induced surface tension gradients $\Delta P_t$ and viscous shear $\Delta P_s$. (d) Schematic of the experimental setup with heating and cooling $Q$ via thermoelectric (Peltier) modules. Scale bars represent (a) 10 mm, (b) 300 $\mu$m, and (c) 200 $\mu$m.}
      \label{Fig1}
      \end{figure}

Microfluidic channels were fabricated out of PDMS and polyurethane and bonded to glass coverslips or slides. An array of small cavities ($d = 100$ $\mu$m wide and 150 $\mu$m deep) lined the two side walls. These cavities were used to trap fluids, i.e., air, silicone oil, or dodecanol, generating fluid-fluid interfaces with the bulk fluid (water). A typical PDMS channel with air trapped in the cavities is shown in \fig{Fig1}a-c.

\begin{figure}
\begin{center}
\includegraphics[width=75mm, keepaspectratio=true]
      {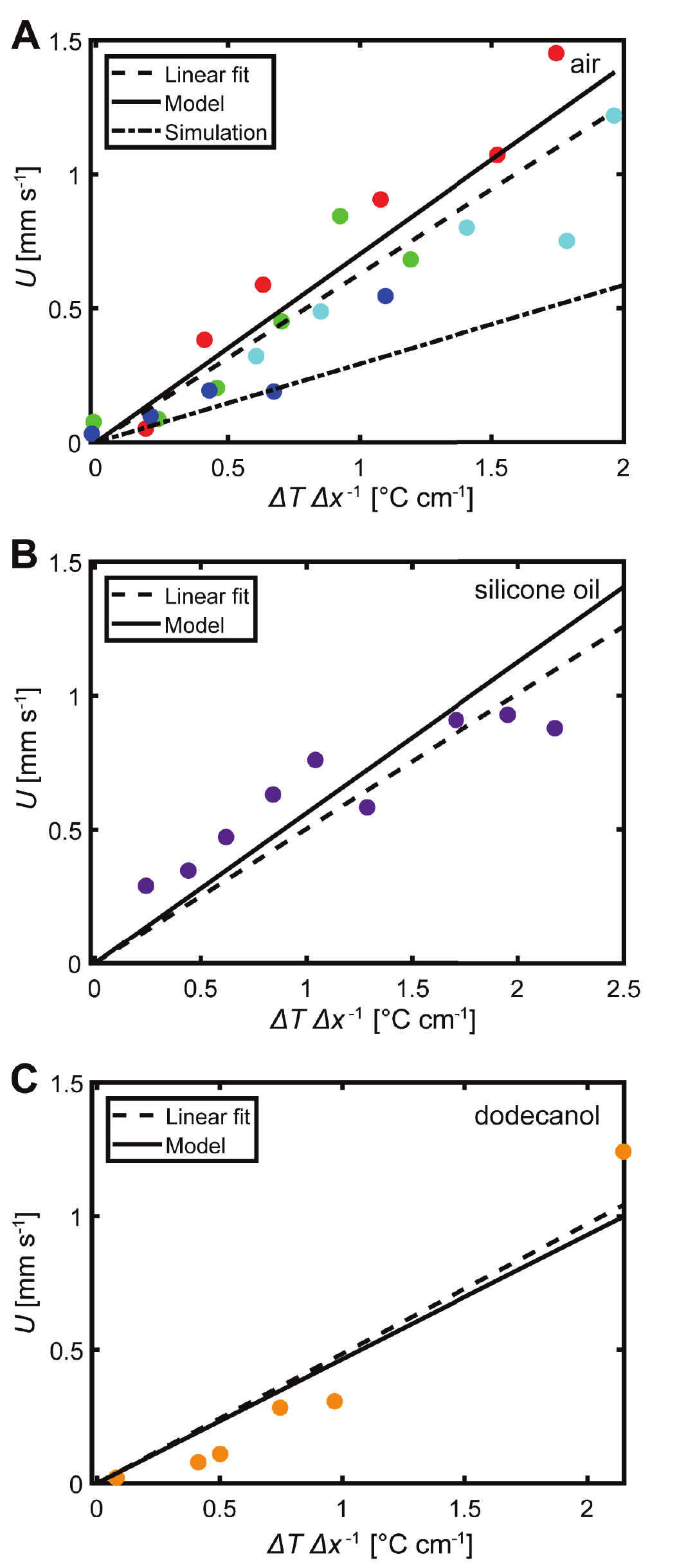}
      \end{center}
      \vspace{0in}
      \caption{{\bf Thermocapillary flow through a microfluidic channel.}
      	(a-c) Relationships between flow speed $U$ and temperature gradient $\Delta T \Delta x^{-1}$ for microchannels with (a) air, (b) silicone oil, and (c) dodecanol filling the cavities along the wall. The dots represent experimental data, with each color corresponding to a different trial. The dashed line represents the best linear fit for the experimental data with (a) $R^2 = 0.84$, (b) $R^2 = 0.52$, and (c) $R^2 = 0.86$. The solid and dash-dot lines represent the fluid mechanics model \eq{Eq1} and numerical simulation, respectively.}
      \label{Fig2}
      \end{figure}

By imposing a temperature gradient across the length of the channel $\Delta T \Delta x^{-1}$, we generated a surface tension gradient that resulted in bulk flow of the water. As shown in \fig{Fig2}, the fluid velocity $U$ along the length of the channel was found to vary linearly with the temperature gradient $\Delta T \Delta x^{-1}$ for all three of the trapped fluids. 

The temperature gradient was generated using two Peltier devices spaced 10 mm apart, as shown in \fig{Fig1}d. An electrical current was applied across each Peltier device to generate heat $Q$ on one of the faces. The Peltier devices were oriented in opposite directions so that one transfers heat $Q$ to the channel, while the other transfers $Q$ out of the channel. The temperature difference $\Delta T$ was measured using two thermocouples placed beneath each Peltier device.

\begin{figure}
	\begin{center}
		\includegraphics[width=175mm, keepaspectratio=true]
		{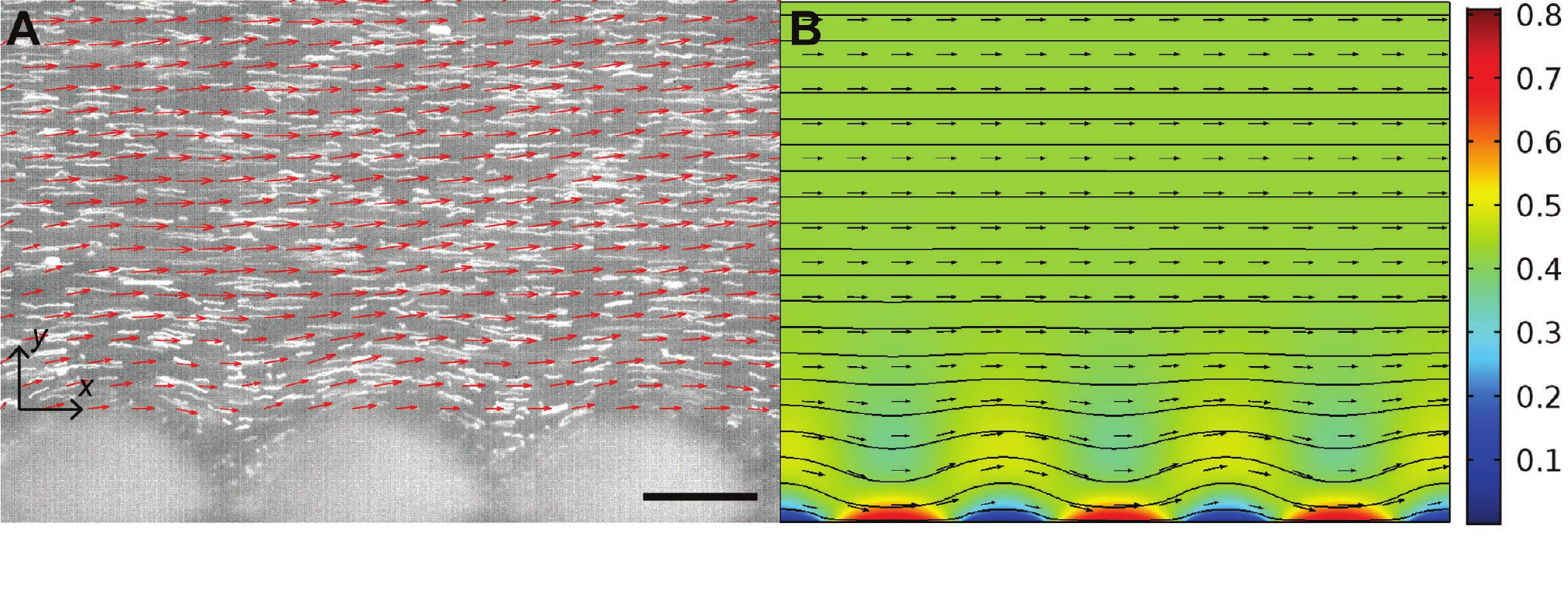}
	\end{center}
	\vspace{0in}
	\caption{{\bf Flow visualization and numerical simulation.} (a) Streaklines and flow field taken from 1 second of video at 200 Hz using micro-PIV. Scale bar represents 50 $\mu$m. (b) Streamlines, flow field, and velocity magnitude (contour in mm s$^{-1}$) from numerical simulation. (a-c) Data are for $\Delta T \Delta x^{-1} = 1.75$ $^{\circ}$C cm$^{-1}$.
	}
	\label{Fig3}
\end{figure}

The fluid velocity $U$ was measured using particle image velocimetry at the micrometer scale (or micro-PIV). Details on the procedure can be found in the Methods. By seeding the bulk water with 1-$\mu$m fluorescent particles and exciting them with a high-powered laser, we observed their motion and deduced the flow within the microfluidic channels. \fig{Fig3}a shows streaklines and the velocity field for a typical experiment with air trapped in the cavities. The streaklines represent the motion of the particles during one second of filming at 200 Hz. Typical videos obtained using micro-PIV are shown in the electronic supplementary material, video S1. By visualizing the flow, we observed bulk fluid motion within the microchannels with a velocity $U$, which was induced by the temperature gradient $\Delta T \Delta x^{-1}$.

The observed linear relationship between $U$ and $\Delta T \Delta x^{-1}$ was modeled through a force balance on the bulk water. Two forces act on the flowing water, viscous and thermocapillary. The viscous forces are $\Delta P_s A_s$, where $\Delta P_s$ is the pressure gradient associated with viscous shear losses and $A_s$ is the surface area on the top and bottom walls of the channel where shearing occurs. The thermocapillary forces are $\Delta P_t A_c$, where $\Delta P_t$ is the pressure gradient associated with thermocapillary effects and $A_c$ is the cross-sectional area of the channel. By using a control volume that encompasses one unit cell of the fluid-fluid interface, we may assume $A_s \approx A_c$ since $h \approx d$, where $h = 75$ $\mu$m is the height of the channel as depicted in \fig{Fig1}c.

%Energy is dissipated from the water through viscous losses and introduced via surface tension gradients from thermocapillary effects, which induce pressure gradients $\Delta P_s$ and $\Delta P_t$, respectively. The two pressure gradients are depicted in \fig{Fig1}c. 

The pressure gradient generated by the temperature-induced surface tension gradient along the fluid-fluid interfaces is $\Delta P_t = \gamma \frac{dT}{dx}$, where $\gamma$ is the temperature dependence of the interfacial tension at the fluid-fluid interface. This value has been previously measured for the fluid-fluid interfaces used in this study \cite{Baier,Peters,Villers}. The values are $\gamma = -0.15$, 12, and 0.10 mN m$^{-1}$ $^{\circ}$C$^{-1}$ for air-water, silicone oil-water, and dodecanol-water, respectively. The interfacial tension between air and water decreases with increasing temperature, while it increases with increasing temperature for water and silicone oil and dodecanol. These values were reported for a temperature range of $25-40$ $^{\circ}$C, and our experiments were conducted within the same range. \fig{Fig1}b-c show how the velocity and temperature gradient are in opposite directions for air-water interfaces. The shear resisting the fluid motion on the top and bottom walls is responsible for pressure losses, expressed as $\Delta P_s = \mu \frac{d{\bf u}}{dz}$, where $\mu$ is the dynamic viscosity of the bulk fluid (water) and ${\bf u}$ is the velocity vector.

The thermocapillary pressure gradient $\Delta P_t$ was simplified by assuming a linear temperature gradient; therefore, $\frac{dT}{dx}=  \Delta T \Delta x^{-1}$, where $\Delta T$ is the temperature difference between the two Peltier devices and $\Delta x$ is the distance between the points where the temperatures were measured. Additionally, the losses due to shear $\Delta P_s$ were simplified by assuming a linear velocity gradient with respect to the channel depth $z$, and uniform velocity across the channel (in $y$). Then, it follows that $\frac{d{\bf u}}{dz} = 2U h^{-1}$. 

With these assumptions and through a force balance, we may solve for the flow velocity $U$ by balancing the two pressure gradients, or $\Delta P_t = \Delta P_s$, thus resulting in

\begin{equation}
U = \frac{\gamma h}{2 \mu} \frac{\Delta T}{\Delta x}.
\label{Eq1}
\end{equation}

We found that this model provided an accurate representation of the observed fluid mechanics. As shown in \fig{Fig2}, the model predicts the slope of the linear relationship within 12\%, 12\%, and 4\% error for air-water, silicone oil-water, and dodecanol-water, respectively.

To gain further insight into this phenomenon, especially the bulk flow, we used numerical simulation. We used a two-dimensional (2D) geometry (see electronic supplementary material \fig{FigS3}) to simulate the flow through the center plane of the channel, where the maximum flow is expected. The boundary conditions used in the simulation (electronic supplementary material \tab{TabS1}) were adapted from a previous study \cite{Baier}. We found that while the numerical simulations captured the linear relationship between $U$ and $\Delta T \Delta x^{-1}$, they predicted much lower velocities (\fig{Fig2}a) with an error of 53\% for the slope.

However, the simulations provided qualitative agreement with the experimental observations (compare \fig{Fig3}a-b). The streaklines and streamlines in the experiments and simulation both exhibited bulk fluid motion even away from the fluid-fluid interfaces along the side walls, as depicted by the velocity distribution in the electronic supplementary material \fig{FigS4}. Therefore, we have successfully demonstrated the first enclosed thermocapillary micro-pump. While others have exploited the thermocapillary phenomenon to drive particles through a  fluid \cite{Maggi,Sammarco,Hu,Ohta} or to transport thin fluid films \cite{Kataoka}, we directly applied thermocapillary stresses to a fluid and generated bulk motion within a closed microchannel. Additionally, we have found that our thermocapillary pump outperforms the one first proposed by previous researchers (compare the dashed and dash-dot lines in \fig{Fig2}a) \cite{Baier}.  

The simple design, a polymer microchannel bonded to glass, and low thermal energy required by the thermocapillary phenomenon, $\sim 1$ watt, make our device attractive for lab-on-a-chip applications, especially in point-of-care (POC). Moderate temperature gradients, on the order of degrees Celsius per centimeter, can pump fluids through microfluidic devices without the use of traditional electrical micro-pumps. Additionally, the phenomenon is not restricted to electrically conductive fluids, like electrocapillary pumping or electro-wetting.

\begin{figure}
	\begin{center}
		\includegraphics[width=175mm, keepaspectratio=true]
		{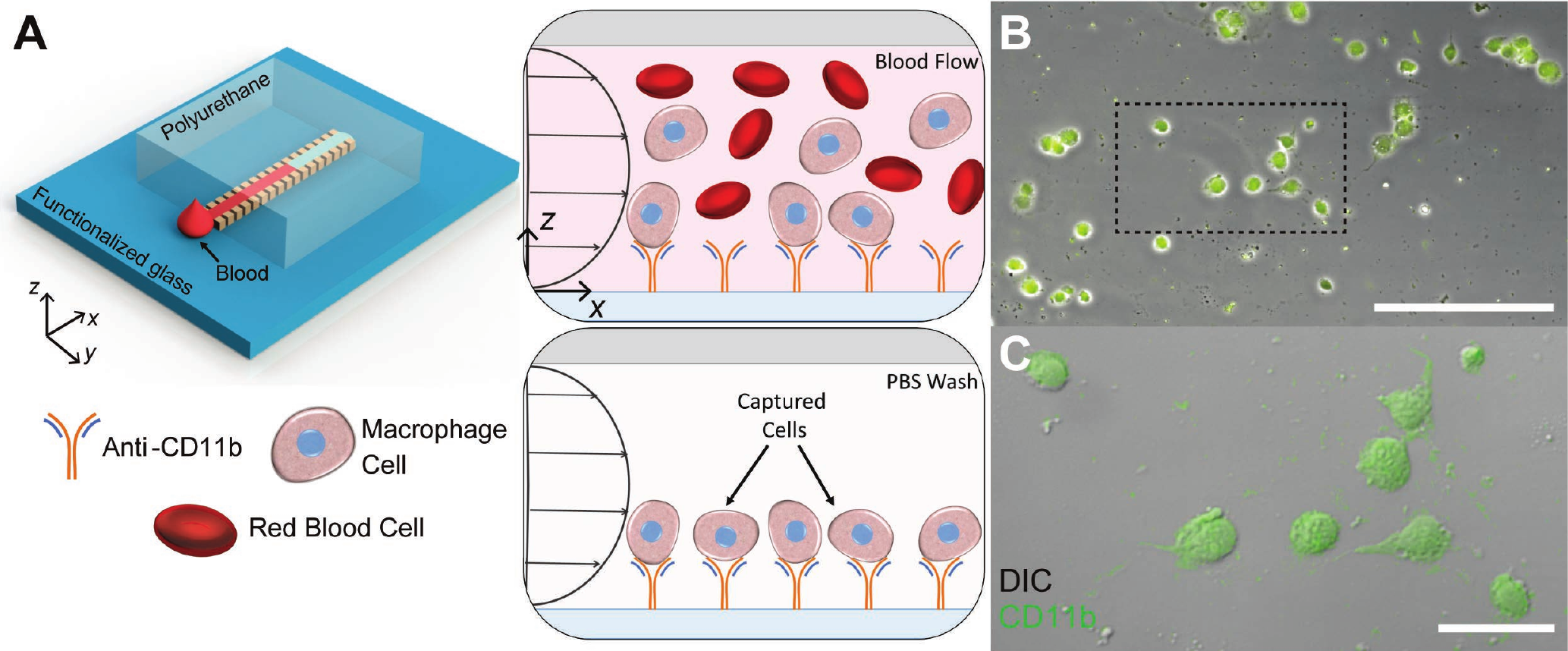}
	\end{center}
	\vspace{0in}
	\caption{{\bf Thermocapillary microfluidic device for counting macrophage cells in blood.} 
	(a) Schematic of the thermocapillary microchannel used to pump blood and capture macrophage cells. The macrophage cells within the blood flow are captured with an anti-body (anti-CD11b). (b-c) Phase and fluorescence (green) images of macrophage cells captured by the anti-bodies on the functionalized glass slide of the microchannel. The dashed rectangle in (b) represents the region depicted in (c). Scale bars represent (b) 100 $\mu$m and (c) 25 $\mu$m.}
	\label{Fig4}
\end{figure}

To demonstrate the device's functionality, we used it to pump blood samples spiked with specific leukocyte subpopulations over functionalized glass for selective isolation of the leukocytes within the blood. A schematic of the demonstration is shown in \fig{Fig4}a. A drop of blood spiked with CD11b+ monocytes and macrophages was placed at the inlet of the microchannel, and then pumped through using thermocapillary effects. After the blood was pumped through, phosphate-buffered saline (PBS) was pumped to wash away red blood cells. The CD11b+ monocytes and macrophages were immobilized on the glass via binding to antibodies functionalized on the surface (Anti-CD11b). Details on the surface functionalization can be found in the electronic supplementary material \fig{FigS5}. With our demonstration, we found effective isolation of CD11b+ leukocytes with high specificity, as shown in the ﬂuorescent images in \fig{Fig4}b-c. Our thermocapillary micropump was capable of pumping a heterogeneous biological fluid and power a functional cell isolation assay, and can be further utilized for diagnostic and monitoring at the POC. 

%To demonstrate the device's functionality, we used it to pump blood flow over functionalized glass and capture and count macrophage cells within the blood. A schematic of the demonstration is shown in \fig{Fig4}a. A drop of blood with macrophage cells was placed at the inlet of the microchannel, and then pumped through using thermocapillary effects. After the blood was pumped through, phosphate-buffered saline (PBS) was pumped to wash the blood away. The macrophage cells were captured on the functionalized glass due to binding to anti-bodies (anti-CD11b). Details on the functionalization can be found in the electronic supplementary material \fig{FigS5}. With our demonstration, we found efficient capture of macrophage cells, as shown in the fluorescent images in \fig{Fig4}b-c. Our thermocapillary micropump was capable of pumping a biological fluid, and may be utilized in diagnostic tools.

\begin{figure}
     	\begin{center}
     		\includegraphics[width=125mm, keepaspectratio=true]
     		{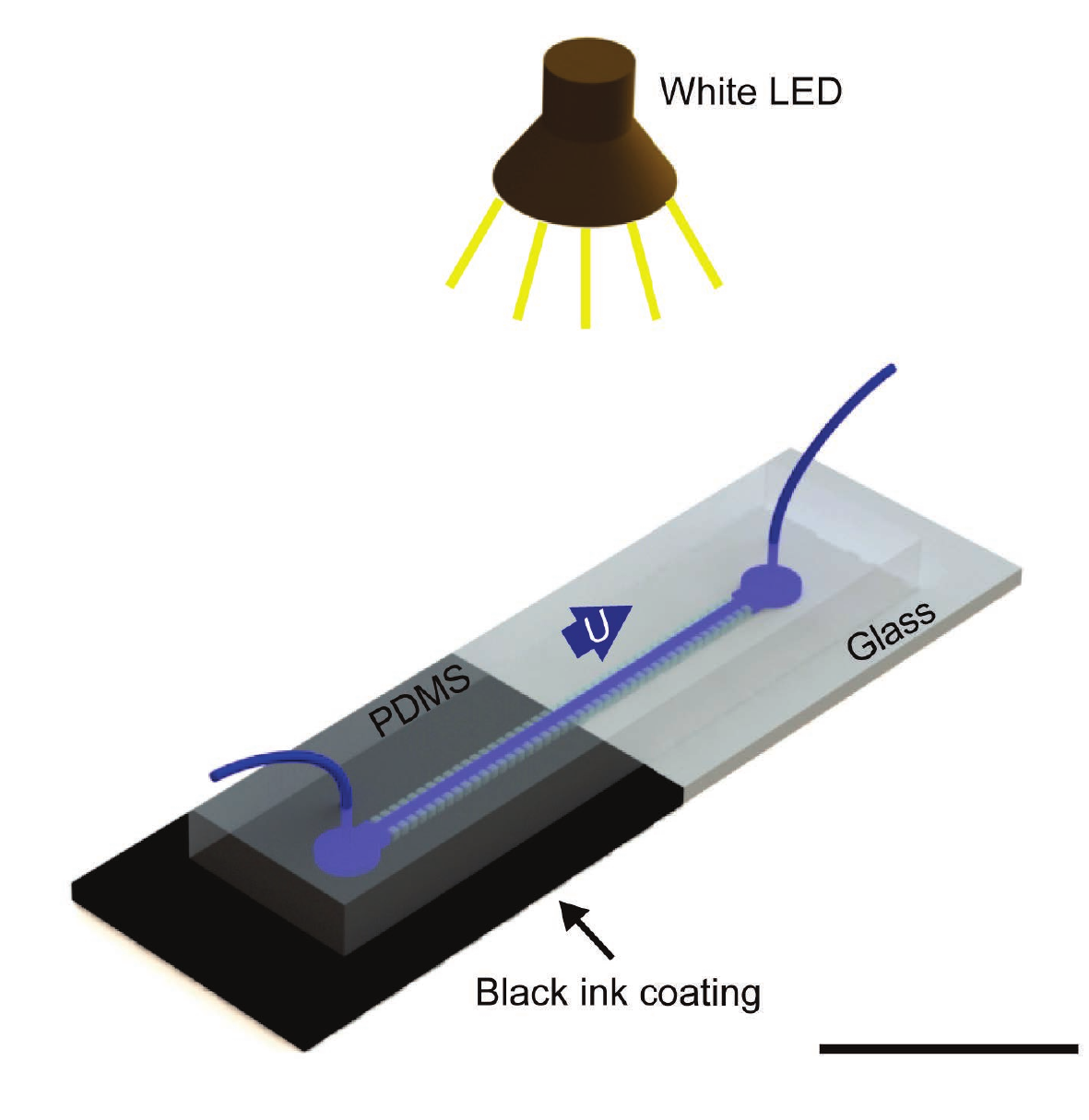}
     	\end{center}
     	\vspace{0in}
     	\caption{{\bf Thermocapillary microfluidic device powered by solar irradiation.} 
     		Schematic of the thermocapillary microchannel powered by solar irradiation simulated by a high-intensity white LED. Scale bar represents 25 mm.}
     	\label{Fig5}
     \end{figure}

We tested to determine if our device could be powered by solar irradiation at the POC.  A schematic of the experiment is shown in \fig{Fig5}. Air was trapped in the cavities by water and the flow was observed with 1-$\mu$m polystyrene seeding particles. A high-intensity white LED was used to simulate the typical irradiation experienced on the Earth's surface, or 100 mW cm$^{-2}$. Half of the thermocapillary channel was coated with black ink to absorb the thermal energy and generate a temperature gradient. The electronic supplementary material, video S2, shows the observed flows for irradiations of 44 and 88 mW cm$^{-2}$, which produced temperature differences $\Delta T = 1$ and $4$ $^{\circ}$C, respectively. The observed flows were significant, or $U = 5$ and 10 $\mu$m s$^{-1}$, respectively. Therefore, solar irradiation provided enough thermal energy to drive fluid flow. Thermocapillary-driven microfluidic devices may be operated in remote locations where electricity is not available.

\section{Discussion}

In this study, we presented the design of an enclosed thermocapillary-driven microfluidic channel. This device exploits surface tension gradients induced by temperature differences for fluid transport. We used micro-PIV to provide a quantitative observation of the fluid flow. Numerical simulation and fluid mechanic modeling revealed the underlying physics. The model provided an accurate representation of the phenomenon; therefore, the channel geometry and working fluids may be designed to generate the desired fluid flows.  

We investigated the thermocapillary flow with three different fluid-fluid interfaces: air-water, silicone oil-water, and dodecanol-water. However, it is not limited to these particular interfaces. Other fluid-fluid interfaces may be used, and future studies should be pursued to determine other successful interfaces. The selection of working fluids may be a function of the application since their interfacial tension will determine the flow direction and velocity. The geometry of the microchannel may also be tuned to create more complex flows, especially with integration of MEMS.

Microfluidics offer the potential to revolutionize clinical practice in diagnosis and monitoring of various debilitating diseases, from cancer to sickle cell disease, due to their time- and cost-efficient operation using only miniscule volume of reagents and clinical samples. Even though most developed microfluidic technologies can fit on tiny cover glasses with small fingerprints, these technologies also require bulky periphery equipment for operation, such as syringe pumps. Use of syringes and syringe pumps increases the required reagent and sample volumes due to dead volumes within the syringe, needle, and long tubing. All these factors significantly hamper the application and translation of microfluidic technologies at the POC, especially at low-resource settings with limited electricity supply in the field. The thermocapillary micropump presented here has the potential to replace bulky syringe pumps and, due to its low power consumption, can enable translation of various microfluidic clinical assays for portable and hand-held operation at the POC.  

In short, this platform is capable of continuously driving fluid flows without the use of conventional fluid pumps and irrespective of the electrical conductivity of the working fluid. We demonstrated our device’s capability of functioning with heterogeneous biological ﬂuids, in this case blood, for diagnostic applications. Additionally, our device was able to pump fluids using only solar irradiation, making it possible to operate without the use of electricity.

\section{Methods}

\subsection{Microfluidic channel fabrication}

The microfluidic channels were fabricated using photolithography and polymer molding. The channels were 27 mm long, 1 mm wide, and $h = 75$ $\mu$m deep with two circular reservoirs (3 mm diameter) for the inlet and outlet. The exact height $h$ of the channels was measured using a laser profilometer (Keyence VK-X200, Neu-Isenburg, Germany), as shown in the electronic supplementary material \fig{FigS1}. The laser scans were conducted using an objective with 100X magnification. The walls of the channel were lined with 150-$\mu$m long and 50-$\mu$m wide fins, equally spaced by $d = 100$ $\mu$m. The space between the fins trapped air, silicone oil, or dodecanol, as shown in \fig{Fig1}a-c.

For fabrication, first, a mask of the channel was fabricated on sodalime glass by Compugraphics Jena GmbH (Jena, Germany). This mask was then used to fabricate a mold using SU-8 photoresist (micro resist technology GmbH, Berlin, Germany). The photoresist was spin-coated (WS-650MZ-8NPPB, Laurell Technologies, North Wales, PA, USA) to a thickness of approximately 100 $\mu$m onto a 3-in silicone wafer. Next, the photoresist was etched using UV irradiation (MJB 4 mask aligner, SUSS MicroTec, Garching, Germany). Finally, the SU-8 mold was used to cast polydimethylsiloxane (PDMS; Sylgard 184, Dow Corning, Inc., Midland, MI, USA) and polyurethane (ST-1087, BJB Enterprises, Inc., Tustin, CA, USA) channels. In order to seal a channel, it was bonded to a glass coverslip (24 x 60 x 0.15 mm). The bonding was achieved by first etching the channel and glass using oxygen plasma (Diener Electronic Zepto, Ebhausen, Germany). The channel and glass were pressed together and then baked for 30 minutes at 60 $^{\circ}$C. This process ensured a tight bond and prevented any leakage.

\subsection{Particle image velocimetry (PIV) and heating and cooling through the Peltier effect}

The fluid flow inside of the microfluidic channels was characterized through the use of particle image velocimetry (PIV). Water was seeded with 1-$\mu$m polystyrene particles embedded with a fluorochrome dye (Molecular Probes, Inc., Eugene, OR, USA; 1.1 $\pm$ 0.035 $\mu$m) that excites at a wavelength of 535 nm and emits at 575 nm for fluorescence imaging. The particles were excited using a 10 mJ, 1000 Hz, 527-nm laser (Dantec Dynamics A/S, Skovlunde, Denmark) and imaged through a fluorescence stereomicroscope (Leica M165 FC with PLAN APO 2.0x CORR objective, Wetzlar, Germany). The videos were captured using a high speed camera (Vision Research SpeedSense M310, Wayne, NJ, USA) at 200 Hz for 500 frames, or 2.5 seconds, at 58x magnification. Typical videos obtained are shown in the electronic supplementary material, video S1.

For processing the images obtained, we used the built-in functions in DynamicStudio 2016a (Dantec Dynamics) software. The same analysis sequence was used for each trial. First, we coupled each image pair, or made double frames. With this function, we were able to obtain 250 image pairs from the 500 frames captured. Next, we defined a mask to exclude the fins and fluid trapped in the cavities along the channel wall, focusing only on water in the bulk. Then, we found the minimum image value for each image pair and subtracted it in order to eliminate the background and isolate the particles. Next, we applied a low-pass Gaussian filter. Then, we again subtracted the minimum image value to further eliminate the background. Finally, we calculated vector fields for each image pair with a vector every 32 pixels. Typical image pairs obtained after each processing step are shown in the electronic supplementary material \fig{FigS2}.

After obtaining the vector fields, we averaged them to obtain an overall average vector field for each trial. To report the flow velocity $U$, we found the average horizontal velocity within a 350 $\mu$m by 20 $\mu$m rectangle immediately above the trapped fluid. This value is the velocity $U$ reported throughout the text and in \fig{Fig2}.

Because of the inlet and outlet conditions of the microchannels, there was an observed minimum temperature gradient $\Delta T$ required to initiate the flow. The inlet and outlets of the channels were connected to tubing (ID 0.5 mm, OD 1.5 mm, length 50 mm; Tygon$^{\tiny{\textregistered}}$ Tubing, Cole-Parmer GmbH, Wertheim, Germany) via an inserted metal tube (24 gauge: ID 0.31 mm, OD 0.57 mm, length 8 mm). This connection introduced a 90$^{\circ}$ turn into the channel, as well as a constriction followed by an expansion in the width of the fluid. Flow constrictions have been observed to cause pressure drops in microchannels in the Stokes flow regime \cite{Lee}, akin to flows within ducts in the laminar flow regime \cite{Munson}. In order to correct for this observed phenomenon, we corrected the temperature gradient $\Delta T$ for each trial so that $U = 0$ at $\Delta T = 0$. Each trial was corrected individually due to slight variations in the fabrication of each channel. This correction was made for all trials (\fig{Fig2}).

The temperature gradient across the microfluidic channels was induced using thermoelectric modules that use the Peltier effect (European Thermodynamics Limited ETC-200-14-06-E, Leicestershire, UK). Two modules were used, one to heat and one to cool, spaced 10 mm apart. They were supplied electrical  currents $i$ varying from 0.8 to 1.6 amps. To ensure good thermal contact between the modules and the channels, we applied silicone thermal grease (Heat Sink Compound Plus, 2.9 W/m-K; RS Components Ltd., Northants, UK) evenly on the contact surface.

The channel and thermoelectric modules were mounted in a custom, 3D-printed stand \fig{Fig2}a. Additionally, thermocouples were integrated to measure the temperature difference $\Delta T$ (OMEGA Engineering RDXL6SD with Type K thermocouples, Norwalk, CT, USA). The heating/cooling, temperature measurement, and PIV recording were synchronized using an Arduino processor (make and model). The experimental procedure is as follows. First, the heating and cooling are initiated. Then, the PIV videos are captured after 30 seconds. Finally, the temperature difference is measured at the instant when acquisition started. The 30 second delay ensures that the temperature difference and flow velocity reach a steady state.

\subsection{Numerical simulation}

We used COMSOL Multiphysics$^{\tiny{\textregistered}}$ Software 5.3 (COMSOL, Inc., Stockholm, Sweden) to conduct numerical simulations of our experimental setup depicted in \fig{Fig1}b. A simplified, two-dimensional representation of the setup was constructed. We simulated the flow over one cavity by exploiting the symmetry and periodic boundary conditions, akin to the simulations reported in previous studies \cite{Baier,Baier01}. A schematic of the geometry and boundary conditions is presented in the electronic supplementary material \fig{FigS3} and \tab{TabS1}. The velocity $U$ reported was the average horizontal velocity within the 100 $\mu$m by 20 $\mu$m rectangle immediately above the air pocket.

The simulation was first verified by comparing it to the results from a previous study for a temperature gradient of 6 $^{\circ}$C cm$^{-1}$ \cite{Baier01}. Next, a mesh study was conducted across two orders of magnitude in mesh elements. The results are shown in the Supplement \tab{TabS2}. They closely match the reported velocity of 1.7 mm s$^{-1}$ from Baier \et \ for $\Delta T \Delta x^{-1} = 6$ $^{\circ}$C cm$^{-1}$ \cite{Baier01}. The simulations shown in \fig{Fig2}a and \fig{Fig3}b were conducted using the highest number of mesh elements.

Finally, the temperature gradients generated during the experiments were simulated. The results were compared to the experiments, shown in \fig{Fig2}a. Streamlines and flow profiles for three fluid cavities were plotted for a qualitative comparison to the experimental observations (\fig{Fig3}b). 

\subsection{Macrophage cell capture using thermocapillary pumping}

Microchannels were bonded with glass slides coated with 3-Aminopropyl Triethoxysilane (Electron Microscopy Sciences, Hatfield, PA) and functionalized against the target cell types using a three-step surface chemistry approach \cite{Alapan,Krebs}. Microfluidic channels were first injected with sulfo-N-g-maleimidobutyryloxysuccinimide ester solution (0.28\% (v/v) in water) for 15 minutes at room temperature. Afterwards, microfluidic channels were flushed with PBS and filled with Neutravidin solution (1 mg mL$^{-1}$ in PBS). After an incubation period of 30 minutes, the channels were washed with PBS to remove any remaining non-bound Neutravidin. Next, biotinylated CD11b antibodies were injected into the channels and incubated for 30 minutes at room temperature. Channels were rinsed again before blood processing. A schematic of the functionalized glass is shown in the electronic supplementary material \fig{FigS5}.
 
Purified mouse RBCs were purchased from Innovative Research (Novi, Michigan). RBCs were washed three times with PBS and reconstituted at a concentration of 108 cells mL$^{-1}$. Mouse monocytes and macrophages (J774A.1, ATCC$^{\tiny{\textregistered}}$ TIB-67TM) were cultured in Dulbecco's Modified Eagle's Medium (DMEM) with 10\% fetal bovine serum and 1\% penicillin-streptomycin (5,000 U mL$^{-1}$). Cultured monocytes and macrophages were scraped and mixed with blood with a final concentration of 106 cells mL$^{-1}$. Afterwards, the diluted blood samples containing monocytes and macrophages were pumped through the channels. Next, PBS was pumped to remove any non-attached cells. Then, channels were injected with fluorescently labeled (Alexa Fluor 488) CD11b antibodies and incubated for 30 minutes at room temperature to determine the specificity of captured cells using a fluorescent microscope (Nikon Eclipse Ti, D{\"u}sseldorf, Germany).

\subsection{Experiments with simulated solar irradiation}

We investigated the potential use of solar irradiation for providing the temperature gradient to drive thermocapillary flow. Solar irradiation was absorbed with black ink (Pelikan 4001 Brilliant Black, Pelikan Vertriebsgesellschaft mbH \& Co, Hannover, Germany). The black ink was applied to the underside of a glass slide (25 x 70 x 1 mm) with a PDMS microchannel bonded to the upper side. Half of the slide was coated with the ink and left to dry.

After the ink dried, the microchannel was filled with water seeded by polystyrene microparticles (Polysciences, Inc., Washington, PA, USA; diameter: $0.99 \pm 0.03$ $\mu$m). The particles allowed approximate observation of fluid motion.

Solar irradiation was simulated using a high-power LED (SOLIS-3C, Thorlabs, Inc., Newton, NJ, USA). The irradiation from the LED was measured with a microscope slide thermal power sensor (S175C, Thorlabs, Inc., Newton, NJ, USA). For the experiments, the LED was mounted to an inverted microscope (Zeiss Axio Observer.A1, Jena, Germany). The microchannel and fluid flow were observed using a 10X objective and high-speed camera (Vision Research Phantom v641, Wayne, NJ, USA). Videos captured at different irradiations are shown in the electronic supplementary material, video S2. The fluid velocities were approximated by tracking three individual particles using an open source tracking software (Tracker by Douglas Brown, http://physlets.org/tracker/).

\section*{Acknowledgments} 
We thank W. Hu for his early contributions, V. Sridhar for help with the solar irradiation experiments, and N. Krishna-Subbaiah for his help with photolithography.

\section*{Author Contributions}
GJA designed the study, carried out the experiments, developed the fluid model, participated in the numerical simulations, and drafted the manuscript; AFT designed the numerical simulations, participated in the design of the study, participated in the experiments, and drafted the manuscript; ZR participated in the experiments and drafted the manuscript; YA designed and participated in the biological experiments, and drafted the manuscript; OY designed and participated in the biological experiments, and drafted the manuscript; MS participated in the design of the study and drafted the manuscript. All authors gave final approval for publication.

\section*{Funding}
This research is funded by the Max Planck Society.

\section*{Data Accessibility}
The datasets supporting this article have been uploaded as part of the supplementary material.

\clearpage
 \bibliographystyle{naturemag}
 \bibliography{thermocapillary}

\begin{thebibliography}{10}
\expandafter\ifx\csname url\endcsname\relax
  \def\url#1{\texttt{#1}}\fi
\expandafter\ifx\csname urlprefix\endcsname\relax\def\urlprefix{URL }\fi
\providecommand{\bibinfo}[2]{#2}
\providecommand{\eprint}[2][]{\url{#2}}

\bibitem{Fan}
\bibinfo{author}{Fan, L.-S.}, \bibinfo{author}{Tai, Y.-C.} \&
  \bibinfo{author}{Muller, R.~S.}
\newblock \bibinfo{title}{Integrated movable micromechanical structures for
  sensors and actuators}.
\newblock \emph{\bibinfo{journal}{IEEE Transactions on Electron Devices}}
  \textbf{\bibinfo{volume}{35}}, \bibinfo{pages}{724--730}
  (\bibinfo{year}{1988}).

\bibitem{Ho}
\bibinfo{author}{Ho, C.-M.} \& \bibinfo{author}{Tai, Y.-C.}
\newblock \bibinfo{title}{{Micro-electro-mechanical-systems (MEMS) and fluid
  flows}}.
\newblock \emph{\bibinfo{journal}{Annual Review of Fluid Mechanics}}
  \textbf{\bibinfo{volume}{30}}, \bibinfo{pages}{579--612}
  (\bibinfo{year}{1998}).

\bibitem{Stone}
\bibinfo{author}{Stone, H.~A.}, \bibinfo{author}{Stroock, A.~D.} \&
  \bibinfo{author}{Ajdari, A.}
\newblock \bibinfo{title}{Engineering flows in small devices: microfluidics
  toward a lab-on-a-chip}.
\newblock \emph{\bibinfo{journal}{Annual Review of Fluid Mechanics}}
  \textbf{\bibinfo{volume}{36}}, \bibinfo{pages}{381--411}
  (\bibinfo{year}{2004}).

\bibitem{Sitti}
\bibinfo{author}{Sitti, M.}
\newblock \emph{\bibinfo{title}{Mobile Microrobotics}} (\bibinfo{publisher}{MIT
  Press}, \bibinfo{address}{Cambridge, MA, USA}, \bibinfo{year}{2017}).

\bibitem{Sengupta}
\bibinfo{author}{Sengupta, S.} \emph{et~al.}
\newblock \bibinfo{title}{Self-powered enzyme micropumps}.
\newblock \emph{\bibinfo{journal}{Nature chemistry}}
  \textbf{\bibinfo{volume}{6}}, \bibinfo{pages}{415--422}
  (\bibinfo{year}{2014}).

\bibitem{Das}
\bibinfo{author}{Das, S.} \emph{et~al.}
\newblock \bibinfo{title}{Harnessing catalytic pumps for directional delivery
  of microparticles in microchambers}.
\newblock \emph{\bibinfo{journal}{Nature Communications}}
  \textbf{\bibinfo{volume}{8}} (\bibinfo{year}{2017}).

\bibitem{Darhuber}
\bibinfo{author}{Darhuber, A.~A.} \& \bibinfo{author}{Troian, S.~M.}
\newblock \bibinfo{title}{Principles of microfluidic actuation by modulation of
  surface stresses}.
\newblock \emph{\bibinfo{journal}{Annual Review of Fluid Mechanics}}
  \textbf{\bibinfo{volume}{37}}, \bibinfo{pages}{425--455}
  (\bibinfo{year}{2005}).

\bibitem{Karbalaei}
\bibinfo{author}{Karbalaei, A.}, \bibinfo{author}{Kumar, R.} \&
  \bibinfo{author}{Cho, H.~J.}
\newblock \bibinfo{title}{Thermocapillarity in microfluidics--a review}.
\newblock \emph{\bibinfo{journal}{Micromachines}} \textbf{\bibinfo{volume}{7}},
  \bibinfo{pages}{13} (\bibinfo{year}{2016}).

\bibitem{Bourouiba}
\bibinfo{author}{Bourouiba, L.}, \bibinfo{author}{Hu, D.~L.} \&
  \bibinfo{author}{Levy, R.}
\newblock \bibinfo{title}{Surface-tension phenomena in organismal biology: An
  introduction to the symposium}.
\newblock \emph{\bibinfo{journal}{American Zoologist}}
  \textbf{\bibinfo{volume}{54}}, \bibinfo{pages}{955--958}
  (\bibinfo{year}{2014}).

\bibitem{Bush}
\bibinfo{author}{Bush, J.~W.} \& \bibinfo{author}{Hu, D.~L.}
\newblock \bibinfo{title}{Walking on water: Biolocomotion at the interface}.
\newblock \emph{\bibinfo{journal}{Annual Review of Fluid Mechanics}}
  \textbf{\bibinfo{volume}{38}}, \bibinfo{pages}{339--369}
  (\bibinfo{year}{2006}).

\bibitem{Money}
\bibinfo{author}{Money, N.~P.}
\newblock \bibinfo{title}{More g's than the space shuttle: Ballistospore
  discharge}.
\newblock \emph{\bibinfo{journal}{Mycologia}} \bibinfo{pages}{547--558}
  (\bibinfo{year}{1998}).

\bibitem{Federle}
\bibinfo{author}{Federle, W.}, \bibinfo{author}{Riehle, M.},
  \bibinfo{author}{Curtis, A.~S.} \& \bibinfo{author}{Full, R.~J.}
\newblock \bibinfo{title}{An integrative study of insect adhesion: mechanics
  and wet adhesion of pretarsal pads in ants}.
\newblock \emph{\bibinfo{journal}{Integrative and Comparative Biology}}
  \textbf{\bibinfo{volume}{42}}, \bibinfo{pages}{1100--1106}
  (\bibinfo{year}{2002}).

\bibitem{Kim}
\bibinfo{author}{Kim, W.}, \bibinfo{author}{Gilet, T.} \&
  \bibinfo{author}{Bush, J.~W.}
\newblock \bibinfo{title}{Optimal concentrations in nectar feeding}.
\newblock \emph{\bibinfo{journal}{Proceedings of the National Academy of
  Sciences}} \textbf{\bibinfo{volume}{108}}, \bibinfo{pages}{16618--16621}
  (\bibinfo{year}{2011}).

\bibitem{Notter}
\bibinfo{author}{Notter, R.~H.}
\newblock \emph{\bibinfo{title}{Lung Surfactants: Basic Science and Clinical
  Applications}} (\bibinfo{publisher}{CRC Press}, \bibinfo{year}{2000}).

\bibitem{Thomson}
\bibinfo{author}{Thomson, J.}
\newblock \bibinfo{title}{{XLII. On certain curious motions observable at the
  surfaces of wine and other alcoholic liquors}}.
\newblock \emph{\bibinfo{journal}{The London, Edinburgh and Dublin
  Philosophical Magazine and Journal of Science.}}
  \textbf{\bibinfo{volume}{10}}, \bibinfo{pages}{330--333}
  (\bibinfo{year}{1855}).

\bibitem{Marangoni}
\bibinfo{author}{Marangoni, C.}
\newblock \bibinfo{title}{Ueber die ausbreitung der tropfen einer
  fl{\"u}ssigkeit auf der oberfl{\"a}che einer anderen}.
\newblock \emph{\bibinfo{journal}{Annalen der Physik}}
  \textbf{\bibinfo{volume}{219}}, \bibinfo{pages}{337--354}
  (\bibinfo{year}{1871}).

\bibitem{Prins}
\bibinfo{author}{Prins, M.}, \bibinfo{author}{Welters, W.} \&
  \bibinfo{author}{Weekamp, J.}
\newblock \bibinfo{title}{Fluid control in multichannel structures by
  electrocapillary pressure}.
\newblock \emph{\bibinfo{journal}{Science}} \textbf{\bibinfo{volume}{291}},
  \bibinfo{pages}{277--280} (\bibinfo{year}{2001}).

\bibitem{Janocha}
\bibinfo{author}{Janocha, B.}, \bibinfo{author}{Bauser, H.},
  \bibinfo{author}{Oehr, C.}, \bibinfo{author}{Brunner, H.} \&
  \bibinfo{author}{G{\"o}pel, W.}
\newblock \bibinfo{title}{Competitive electrowetting of polymer surfaces by
  water and decane}.
\newblock \emph{\bibinfo{journal}{Langmuir}} \textbf{\bibinfo{volume}{16}},
  \bibinfo{pages}{3349--3354} (\bibinfo{year}{2000}).

\bibitem{Chung}
\bibinfo{author}{Chung, S.~K.}, \bibinfo{author}{Ryu, K.} \&
  \bibinfo{author}{Cho, S.~K.}
\newblock \bibinfo{title}{Electrowetting propulsion of water-floating objects}.
\newblock \emph{\bibinfo{journal}{Applied Physics Letters}}
  \textbf{\bibinfo{volume}{95}}, \bibinfo{pages}{014107}
  (\bibinfo{year}{2009}).

\bibitem{Luo}
\bibinfo{author}{Luo, J.} \emph{et~al.}
\newblock \bibinfo{title}{Moving-part-free microfluidic systems for
  lab-on-a-chip}.
\newblock \emph{\bibinfo{journal}{Journal of Micromechanics and
  Microengineering}} \textbf{\bibinfo{volume}{19}}, \bibinfo{pages}{054001}
  (\bibinfo{year}{2009}).

\bibitem{Colgate}
\bibinfo{author}{Colgate, E.} \& \bibinfo{author}{Matsumoto, H.}
\newblock \bibinfo{title}{An investigation of electrowetting-based
  microactuation}.
\newblock \emph{\bibinfo{journal}{Journal of Vacuum Science \& Technology A:
  Vacuum, Surfaces, and Films}} \textbf{\bibinfo{volume}{8}},
  \bibinfo{pages}{3625--3633} (\bibinfo{year}{1990}).

\bibitem{Benard}
\bibinfo{author}{B{\'e}nard, H.}
\newblock \emph{\bibinfo{title}{Les tourbillons cellulaires dans une nappe
  liquide propageant de la chaleur par convection: en r{\'e}gime permanent}}
  (\bibinfo{publisher}{Gauthier-Villars}, \bibinfo{address}{Paris, France},
  \bibinfo{year}{1901}).

\bibitem{Shereshefsky}
\bibinfo{author}{Shereshefsky, J.}
\newblock \bibinfo{title}{{Surface tension of saturated vapors and the equation
  of E{\"o}tv{\"o}s}}.
\newblock \emph{\bibinfo{journal}{The Journal of Physical Chemistry}}
  \textbf{\bibinfo{volume}{35}}, \bibinfo{pages}{1712--1720}
  (\bibinfo{year}{1931}).

\bibitem{Kataoka}
\bibinfo{author}{Kataoka, D.~E.} \& \bibinfo{author}{Troian, S.~M.}
\newblock \bibinfo{title}{Patterning liquid flow on the microscopic scale}.
\newblock \emph{\bibinfo{journal}{Nature}} \textbf{\bibinfo{volume}{402}},
  \bibinfo{pages}{794} (\bibinfo{year}{1999}).

\bibitem{Maggi}
\bibinfo{author}{Maggi, C.}, \bibinfo{author}{Saglimbeni, F.},
  \bibinfo{author}{Dipalo, M.}, \bibinfo{author}{De~Angelis, F.} \&
  \bibinfo{author}{Di~Leonardo, R.}
\newblock \bibinfo{title}{Micromotors with asymmetric shape that efficiently
  convert light into work by thermocapillary effects}.
\newblock \emph{\bibinfo{journal}{Nature Communications}}
  \textbf{\bibinfo{volume}{6}} (\bibinfo{year}{2015}).

\bibitem{Sammarco}
\bibinfo{author}{Sammarco, T.~S.} \& \bibinfo{author}{Burns, M.~A.}
\newblock \bibinfo{title}{Thermocapillary pumping of discrete drops in
  microfabricated analysis devices}.
\newblock \emph{\bibinfo{journal}{AIChE Journal}}
  \textbf{\bibinfo{volume}{45}}, \bibinfo{pages}{350--366}
  (\bibinfo{year}{1999}).

\bibitem{Ohta}
\bibinfo{author}{Ohta, A.~T.}, \bibinfo{author}{Jamshidi, A.},
  \bibinfo{author}{Valley, J.~K.}, \bibinfo{author}{Hsu, H.-Y.} \&
  \bibinfo{author}{Wu, M.~C.}
\newblock \bibinfo{title}{Optically actuated thermocapillary movement of gas
  bubbles on an absorbing substrate}.
\newblock \emph{\bibinfo{journal}{Applied Physics Letters}}
  \textbf{\bibinfo{volume}{91}}, \bibinfo{pages}{074103}
  (\bibinfo{year}{2007}).

\bibitem{Hu}
\bibinfo{author}{Hu, W.}, \bibinfo{author}{Fan, Q.} \& \bibinfo{author}{Ohta,
  A.~T.}
\newblock \bibinfo{title}{An opto-thermocapillary cell micromanipulator}.
\newblock \emph{\bibinfo{journal}{Lab on a Chip}}
  \textbf{\bibinfo{volume}{13}}, \bibinfo{pages}{2285--2291}
  (\bibinfo{year}{2013}).

\bibitem{Baier}
\bibinfo{author}{Baier, T.}, \bibinfo{author}{Steffes, C.} \&
  \bibinfo{author}{Hardt, S.}
\newblock \bibinfo{title}{Thermocapillary flow on superhydrophobic surfaces}.
\newblock \emph{\bibinfo{journal}{Physical Review E}}
  \textbf{\bibinfo{volume}{82}}, \bibinfo{pages}{037301}
  (\bibinfo{year}{2010}).

\bibitem{Peters}
\bibinfo{author}{Peters, F.} \& \bibinfo{author}{Arabali, D.}
\newblock \bibinfo{title}{Interfacial tension between oil and water measured
  with a modified contour method}.
\newblock \emph{\bibinfo{journal}{Colloids and Surfaces A: Physicochemical and
  Engineering Aspects}} \textbf{\bibinfo{volume}{426}}, \bibinfo{pages}{1--5}
  (\bibinfo{year}{2013}).

\bibitem{Villers}
\bibinfo{author}{Villers, D.} \& \bibinfo{author}{Platten, J.}
\newblock \bibinfo{title}{Temperature dependence of the interfacial tension
  between water and long-chain alcohols}.
\newblock \emph{\bibinfo{journal}{The Journal of Physical Chemistry}}
  \textbf{\bibinfo{volume}{92}}, \bibinfo{pages}{4023--4024}
  (\bibinfo{year}{1988}).

\bibitem{Lee}
\bibinfo{author}{Lee, W.~Y.}, \bibinfo{author}{Wong, M.} \&
  \bibinfo{author}{Zohar, Y.}
\newblock \bibinfo{title}{Pressure loss in constriction microchannels}.
\newblock \emph{\bibinfo{journal}{Journal of Microelectromechanical Systems}}
  \textbf{\bibinfo{volume}{11}}, \bibinfo{pages}{236--244}
  (\bibinfo{year}{2002}).

\bibitem{Munson}
\bibinfo{author}{Munson, B.~R.}, \bibinfo{author}{Young, D.~F.},
  \bibinfo{author}{Okiishi, T.~H.} \& \bibinfo{author}{Huebsch, W.~W.}
\newblock \emph{\bibinfo{title}{Fundamentals of Fluid Mechanics, 6th edition}}
  (\bibinfo{publisher}{Wiley}, \bibinfo{year}{2009}).

\bibitem{Baier01}
\bibinfo{author}{Baier, T.}, \bibinfo{author}{Steffes, C.} \&
  \bibinfo{author}{Hardt, S.}
\newblock \bibinfo{title}{Numerical modelling of thermocapillary flow on
  superhydrophobic surfaces}.
\newblock In \emph{\bibinfo{booktitle}{14th International Conference on
  Miniaturized Systems for Chemistry and Life Sciences. Chemical and Biological
  Microsystems Society}} (\bibinfo{year}{2010}).

\bibitem{Alapan}
\bibinfo{author}{Alapan, Y.}, \bibinfo{author}{Hasan, M.~N.},
  \bibinfo{author}{Shen, R.} \& \bibinfo{author}{Gurkan, U.~A.}
\newblock \bibinfo{title}{Three-dimensional printing based hybrid manufacturing
  of microfluidic devices}.
\newblock \emph{\bibinfo{journal}{Journal of Nanotechnology in Engineering and
  Medicine}} \textbf{\bibinfo{volume}{6}}, \bibinfo{pages}{021007}
  (\bibinfo{year}{2015}).

\bibitem{Krebs}
\bibinfo{author}{Krebs, J.~C.}, \bibinfo{author}{Alapan, Y.},
  \bibinfo{author}{Dennstedt, B.~A.}, \bibinfo{author}{Wera, G.~D.} \&
  \bibinfo{author}{Gurkan, U.~A.}
\newblock \bibinfo{title}{Microfluidic processing of synovial fluid for
  cytological analysis}.
\newblock \emph{\bibinfo{journal}{Biomedical Microdevices}}
  \textbf{\bibinfo{volume}{19}}, \bibinfo{pages}{20} (\bibinfo{year}{2017}).

\end{thebibliography}
 
\clearpage

\clearpage

% \begin{table}
% 	\begin{center}
% 		\includegraphics[width=150mm, keepaspectratio=true]
% 		{Figures/fits_table_02}
% 	\end{center}
% 	\vspace{0in}
% 	\caption{{\bf Exponential fits for number of beads removed per step.}}
% 	\label{fits01}
% \end{table}

\renewcommand{\thefigure}{S\arabic{figure}}

\setcounter{figure}{0}

\renewcommand{\thetable}{S\arabic{table}}

\setcounter{table}{0}

       \begin{figure}
 	\begin{center}
 		\includegraphics[width=125mm, keepaspectratio=true]
 		{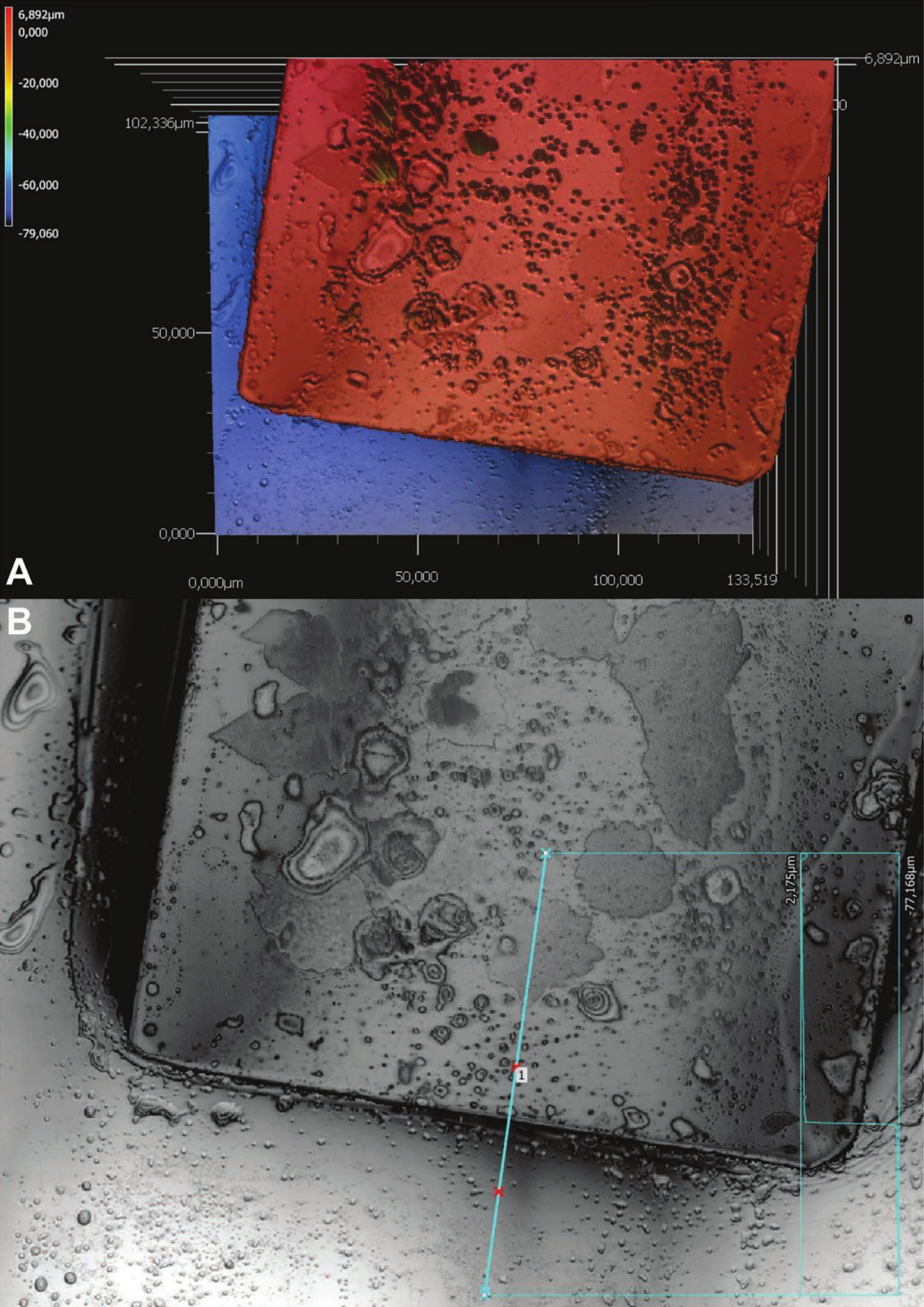}
 	\end{center}
 	\vspace{0in}
 	\caption{{\bf Microchannel height.} 
 		(a-b) Images and measurements taken using a laser profilometer of the SU-8 mold used to cast the microchannels.}
 	\label{FigS1}
 \end{figure}  
 
      \begin{figure}
	\begin{center}
		\includegraphics[width=175mm, keepaspectratio=true]
		{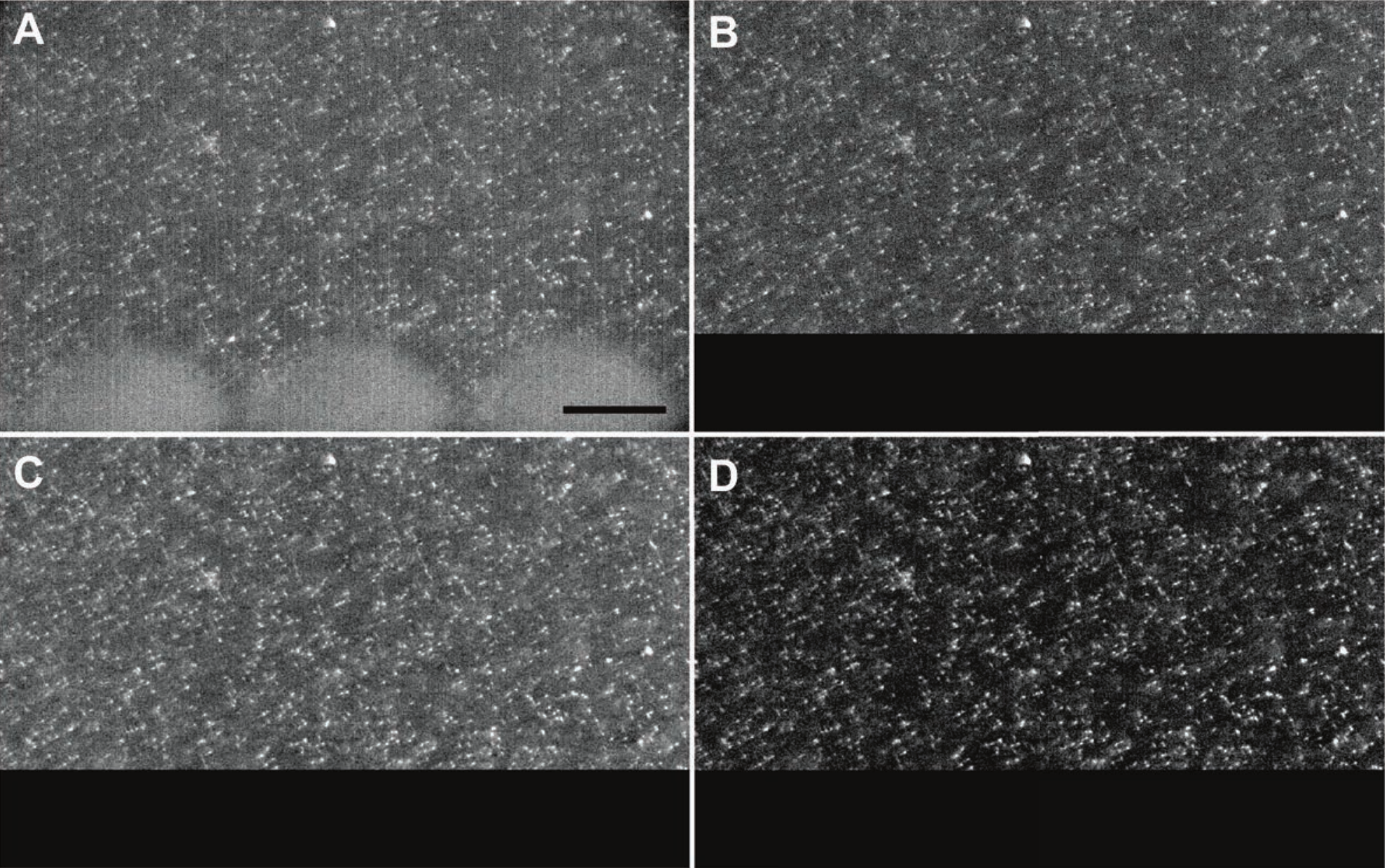}
	\end{center}
	\vspace{0in}
	\caption{{\bf Image processing for PIV.} 
		Typical image pairs obtained from PIV (a) before processing, (b) after subtracting minimum, or background, (c) after applying low-pass Gauss filter, and (d) after subtracting minimum again. The image in (d) is the final image used to obtain vector fields and velocity data. The black rectangles represent the masked area. Scale bar represents 50 $\mu$m.}
	\label{FigS2}
\end{figure}   

      \begin{figure}
	\begin{center}
		\includegraphics[width=100mm, keepaspectratio=true]
		{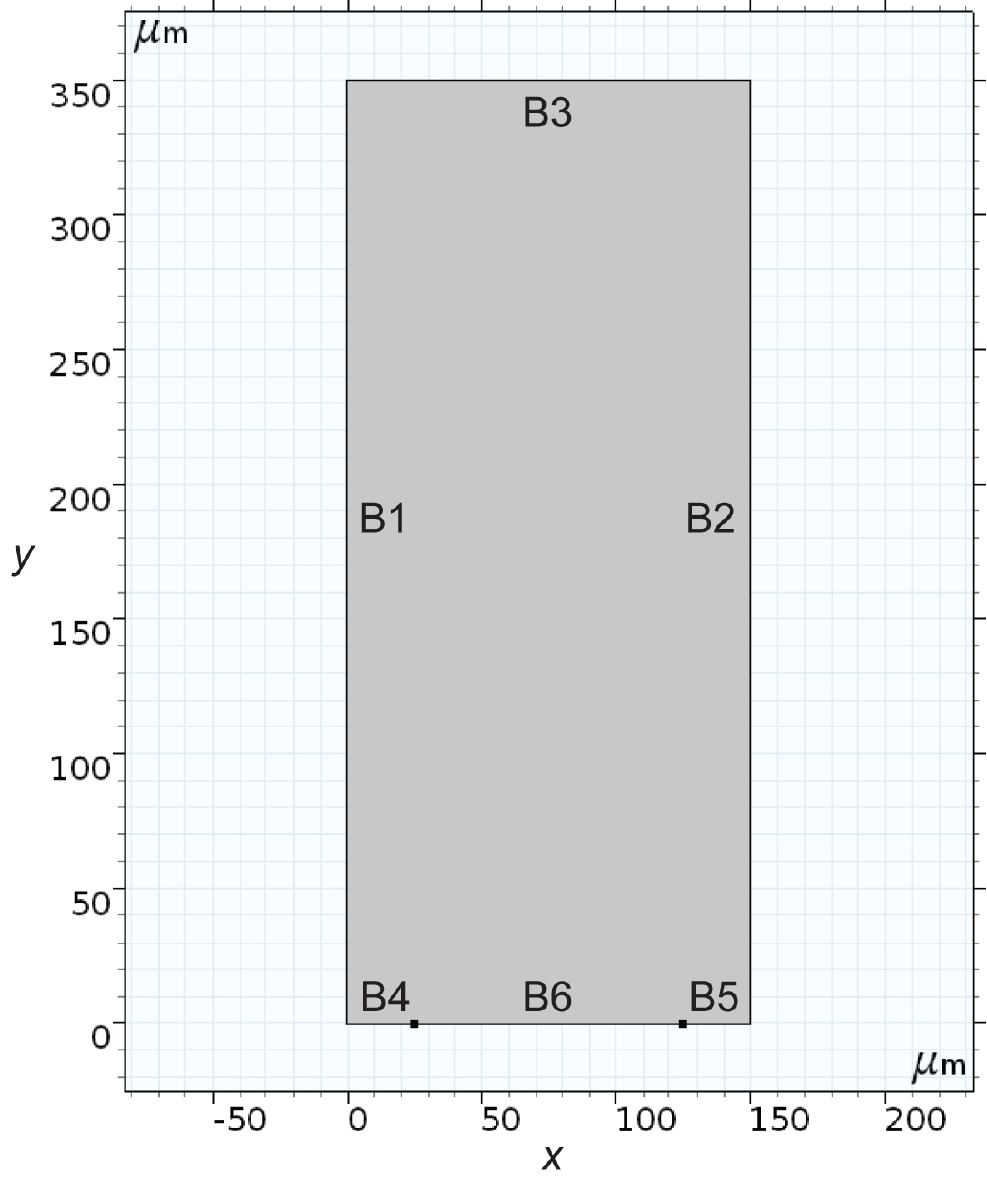}
	\end{center}
	\vspace{0in}
	\caption{{\bf Geometry for numerical simulation.} 
		Schematic depicting the geometry used in COMSOL Multiphysics$^{\tiny{\textregistered}}$ Software. The boundaries are labeled as B1, B2, B3, B4, B5, and B6. The boundary conditions are tabulated in \tab{TabS1}.}
	\label{FigS3}
\end{figure}

\clearpage

 \begin{table}
 	\begin{center}
 		\includegraphics[width=175mm, keepaspectratio=true]
 		{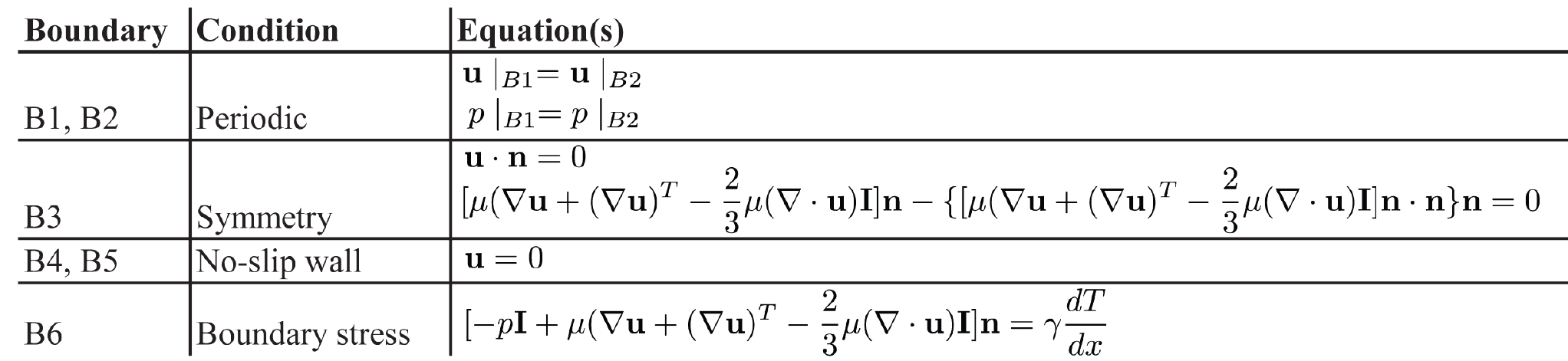}
 	\end{center}
 	\vspace{0in}
 	\caption{{\bf Boundary conditions for numerical simulation.}}
 	\label{TabS1}
 \end{table}    

\begin{table}
	\begin{center}
		\includegraphics[width=125mm, keepaspectratio=true]
		{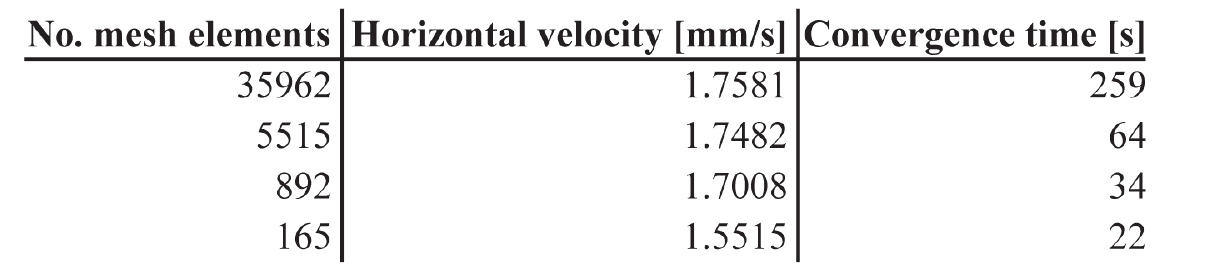}
	\end{center}
	\vspace{0in}
	\caption{{\bf Mesh study for numerical simulation with $\Delta T \Delta x^{-1} = 6$ $^{\circ}$C cm$^{-1}$.}}
	\label{TabS2}
\end{table} 

\clearpage

      \begin{figure}
	\begin{center}
		\includegraphics[width=100mm, keepaspectratio=true]
		{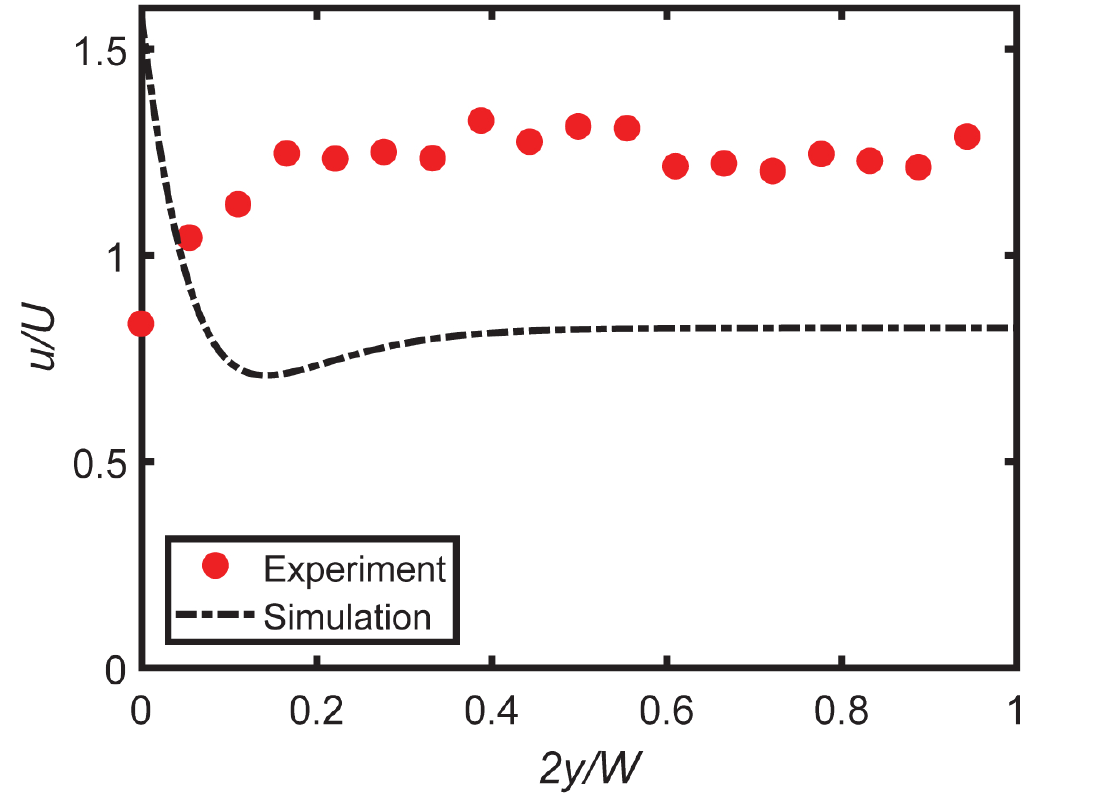}
	\end{center}
	\vspace{0in}
	\caption{{\bf Velocity distribution across channel width.} Relationship between horizontal velocity $u$ and width $y$ of the channel. The velocity is normalized by the measured velocity near the interface $U$, and the width is normalized by half of the total channel width $W$. The results are for $\Delta T \Delta x^{-1} = 1.75$ $^{\circ}$C cm$^{-1}$.
	}
	\label{FigS4}
\end{figure}

\clearpage

      \begin{figure}
	\begin{center}
		\includegraphics[width=150mm, keepaspectratio=true]
		{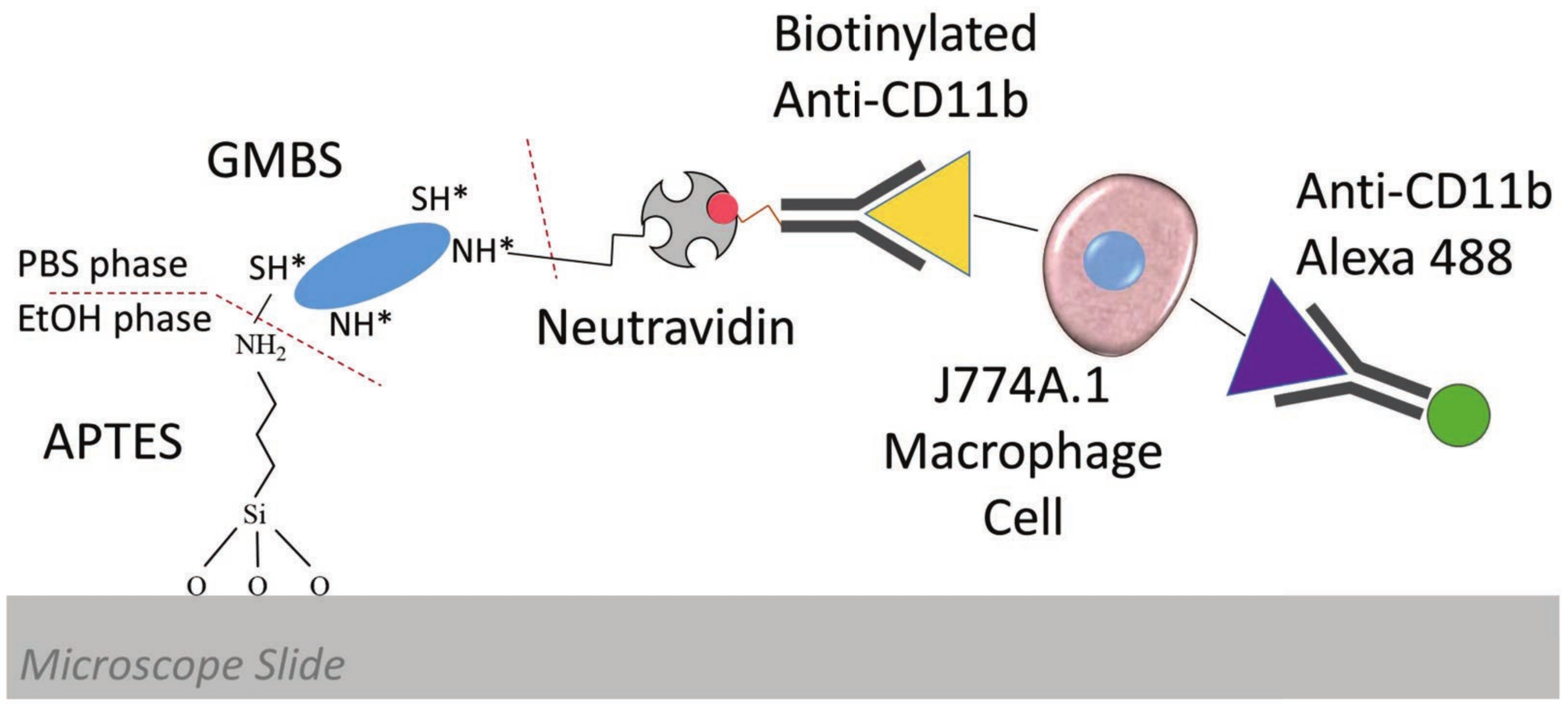}
	\end{center}
	\vspace{0in}
	\caption{{\bf Surface chemistry for macrophage cell capture.} 
		Schematic depicting glass surface treatment with (3-Aminopropyl)triethoxysilane (APTES), N-$\gamma$-maleimidobutyryl-oxysuccinimide ester (GMBS), neutravidin, and anti-body (anti-CD11b) in order to capture macrophage cells in blood flow. The anti-body used is expressed in macrophage cells but not red blood cells.}
	\label{FigS5}
\end{figure}

\end{document}